\documentclass[prl,twocolumn,showpacs,superscriptaddress,floatfix]{revtex4}
\usepackage{graphicx}
\usepackage{bm}
\usepackage{dcolumn}
\usepackage[usenames]{color}
\begin{document}
\newcommand{\bz}{C$_{6}$H$_{6}$}
\newcommand{\Py}{C$_{5}$H$_{5}$N}
%
\title{Chemical versus van der Waals Interaction: The Role of the Heteroatom}
\author{N. Atodiresei}\email{n.atodiresei@fz-juelich.de}
\affiliation{Institut f\"ur Festk\"orperforschung (IFF),
Forschungszentrum J\"ulich, 52425 J\"ulich, Germany}
\affiliation{The Institute of Scientific and Industrial Research,
Osaka University, 8-1 Mihogaoka, Ibaraki Osaka, 567-0047 Japan}
\author{V.~Caciuc}
\affiliation{Institut f\"ur Festk\"orperforschung (IFF),
Forschungszentrum J\"ulich, 52425 J\"ulich, Germany}
\author{P.~Lazi{\'c}}
\affiliation{Institut f\"ur Festk\"orperforschung (IFF),
Forschungszentrum J\"ulich, 52425 J\"ulich,  Germany}
\affiliation{Rudjer Bo\v{s}kovi\'{c} Institute, Zagreb, Croatia.}
\author{S.~Bl\"ugel}
\affiliation{Institut f\"ur Festk\"orperforschung (IFF),
Forschungszentrum J\"ulich, 52425 J\"ulich,  Germany}
\date{\today}
\vspace{-0.4cm}\begin{abstract}
We performed first-principles calculations aimed to investigate the role of an
heteroatom like N in the chemical and the long-range van der Waals (vdW)
interactions for a flat adsorption of several $\pi-$conjugated molecules on
the Cu(110) surface. Our study reveals that the alignment of the molecular
orbitals at adsorbate-substrate interface depends on the number of heteroatoms.
As a direct consequence, the molecule-surface vdW interactions involve not only
$\pi$-like orbitals which are perpendicular to the molecular plane but also
$\sigma$-like orbitals delocalized in the molecular plane.
\end{abstract}
\maketitle
In the last decade, the molecular electronics emerged as a promising field
able to provide the technology necessary to develop devices such as organic
light-emitting diodes (OLEDs) \cite{APL51_913,Shinar_bookOLED}, organic
field-effect transistors \cite{Science283_822,AdvMat14_99,Science303_1644} or
ultra-high-density memory circuits \cite{Nature445_414}. In this context, the
ability to reliably describe the electronic properties
\cite{PRL96_196806,NL7_932} or the mechanical manipulation process of
molecules \cite{Atodiresei2008_1,Atodiresei2008_2} on surfaces is essential
to understand and design the functionality of such devices. The achievement of
these goals strongly depends on the accuracy of the state-of-the-art theoretical
methods used to assess the interaction between a molecule or a molecular layer
and a substrate of choice. Nowadays, the density functional theory (DFT) is the
theoretical tool of choice to analyze
\cite{Atodiresei2007_3,Atodiresei2007_4,PRL96_196806,NL7_932} and to predict
\cite{Atodiresei2007_5,Atodiresei2008_6} the electronic properties of systems
characterized by strong chemical bonds.

However, the loosely bounded physical systems \cite{Bruch2007} represent one
of the major challenges for DFT because the effective Kohn-Sham (KS) potential
does not exhibit the correct asymptotic behavior. In particular, the currently
used exchange correlation energy functionals like the local density
approximation (LDA) or the generalized gradient approximation (GGA) do not
properly describe the long-range van der Waals (vdW) interactions. For
instance, the GGA fails to predict a bonding ground-state for a van der Waals
system like graphite \cite{Carbon44_231}.

One way to circumvent this limit is to include the dispersion effects in DFT
in a semiempirical fashion, in which the total energy of the physical system
is a sum of the self-consistent Kohn-Sham energy as obtained from the DFT
calculations and semiempirical dispersion correction which depends on the
interatomic distance. For example, a semiempirical treatment of the vdW
dispersions in \textit{ab initio} calculations was used to investigate, for
instance, the adenine on graphite(0001) \cite{PRL95_186101}.

Another way to accurately account for the vdW interactions is a first-principle
approach \cite{PRL92_246401} which requires the calculation of a nonlocal
correlation energy functional as a post GGA perturbation procedure. This 
method was successfully employed to investigate the benzene, naphthalene 
and phenol on graphite(0001) \cite{PRL96_146107} and $\alpha$-Al$_2$O$_3$(0001)
or thiophene on Cu(110) \cite{PRL99_176401} surfaces.

In this Letter, we investigate the role of the heteroatoms on the chemical and
the van der Waals interactions for a flat adsorption geometry on the Cu(110)
surface of three prototype $\pi-$electron systems as benzene (Bz), pyridine
(Py) and pyrazine (Pz) molecules. Our \textit{ab initio} simulations reveal
that, for the Bz the long-range dispersion effects are basically important
only for the adsorption energy. This is not the case of the Py and Pz, where
due to their low lying $\pi-$orbitals, the inclusion of the long-range
correlations drastically influence the adsorption geometry and electronic
structure, i. e. the Py becomes chemisorbed on the surface while the Pz binds
to the surface mostly due to the van der Waals interaction. To evaluate the
dispersion effects we used the \emph{ab initio} (vdW-DF)\cite{PRL92_246401} as
well as a semiempirical (DFT-D) \cite{PRB73_205101,JCC27_1787} method.
In particular, a key result of our study is that a state of
the art investigation of the bonding mechanism leading to a flat
molecule$-$substrate adsorption geometry requires the use of both methods to
correctly describe the adsorption-geometry, the electronic structure and the 
correlation effects of the molecule$-$surface interface. 
Moreover, we show that the nature of the
molecular orbitals involved in the dispersion interactions depends on the
alignment of these orbitals with respect to the Fermi level of the
adsorbate$-$surface system.

\begin{figure}[htb]
    {\includegraphics[scale=0.26]{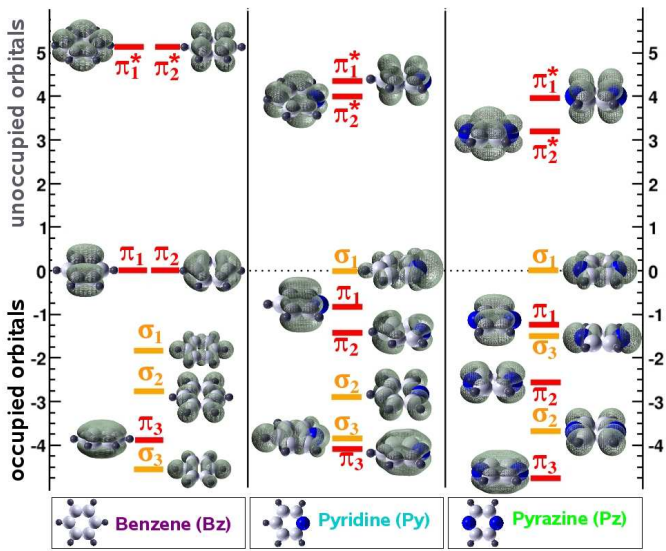}} \\
\caption{(Color online) Energetic levels diagram for Bz 
(left), Py (middle) and Pz (right) molecules. Due to the presence of $N$ atoms 
in the Py and Pz molecules the $\pi-$orbitals are shifted to lower energies 
while some of $\sigma-$orbitals are pushed to higher
energies~\cite{Zeiss1971}.}
\label{fig:DosIsol}
\end{figure}
Our first$-$principles total-energy calculations are carried out in the
framework of density functional theory (DFT) by employing the generalized
gradient approximation (GGA) in a pseudopotential plane-wave formulation
(projector augmented wave method) \cite{PRB50_17953} as implemented in the
VASP code \cite{Kresse1994,Kresse1996}. In particular, in our study we
employed the PBE \cite{PRL77_3865} flavor of GGA.
The plane$-$wave basis set consists of all plane waves up to a kinetic energy
of 450\,eV. The molecule$-$Cu(110) system is modeled within the supercell
approach and contains five atomic layers of copper with the adsorbed molecule
on one side of the slab \cite{Makov1995}. It was generated with the theoretical
bulk copper lattice parameter of 3.63\,{\AA} and has a p(4$\times$6) in-plane
surface unit cell. During our \textit{ab initio} calculations, the uppermost
two copper layers and the molecule atoms were allowed to relax until the
atomic forces are lower than 0.001\,eV/{\AA}.

The evaluation of the vdW forces by using the DFT-D 
\cite{PRB73_205101,JCC27_1787} method in the self-consistent cycle of our
\emph{ab initio} calculations are mandatory to obtain the proper equilibrium
adsorption geometry and the corresponding electronic structure of the Py-- and
Pz--Cu(110) systems. Moreover, we also evaluated the correlation effects by
using the vdW-DF functional \cite{PRL92_246401} as a post processing procedure
(i.e., non-self-consistent) 
on a charge density obtained from the standard GGA calculations 
\cite{Thonhauser2007} using the \cite{JuNoLo} code developed in our group. 

As compared with Bz (C$_6$H$_6$), in the Py (C$_5$H$_5$N) molecule a $CH$ group
of atoms is replaced by a $N$ atom, while the Pz (C$_4$H$_4$N$_2$) has two 
$CH$ groups replaced by two $N$ atoms. All molecules contain a $6\pi-$electron
system and they can adsorb with the molecular plane parallel to the Cu(110)
surface \cite{Komeda2004,Rogers2004,JPCB110_11991}. From electronic point of
view, it is well known \cite{Fle78,Zeiss1971} that the presence of $N$ atoms 
in the conjugated heterocycle molecules lower their $\pi-$orbitals while 
some of $\sigma$ orbitals are pushed to higher energies. This peculiar feature
is clearly shown in the energy levels diagram of the isolated Bz, Py, Pz
molecules (see Fig.~\ref{fig:DosIsol}). 

The Bz is considered to be chemisorbed on the Cu(110) surface
\cite{Komeda2004,Rogers2004} due to an effective hybridization of the
$p_z-$atomic orbitals (those perpendicular to the molecular plane) with the
Cu $d-$bands. Note that these $p_z-$atomic orbitals are forming the
$\pi-$orbitals of the isolated molecule. On the contrary, in the case of Py
and Pz molecules the presence of $N$ atoms lower their $\pi-$orbitals which
implies that the interaction of the corresponding $p_z-$atomic orbitals with
the Cu $d-$bands is expected to be much weaker \cite{Fle78}. Indeed, using
only the standard GGA approximations(PBE) in our DFT calculations of
the flat adsorption geometry of Py and Pz molecules on Cu(110), we find that
the adsorption site is not sensitive to the underlying geometry of the surface.
The average Py(Pz)$-$surface metal distance is 2.98(2.99)\,{\AA}, while for
the Bz this distance is 2.43\,{\AA}. The magnitude of Py-- and Pz--surface
separation distances suggest that these molecules are physisorbed on Cu(110),
while as already mentioned, the Bz is chemisorbed.

\begin{figure}[htb]
    \vspace{-0.4cm}{\includegraphics[scale=0.25]{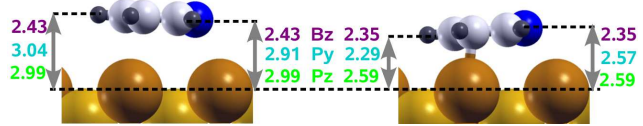}} \\
\caption{(Color online) Side views of the relaxed geometries when
using standard DFT (left panel) and DFT+D (right panel). The inclusion of the
van der Waals interactions at a semi-empirical level in the DFT+D
approach \cite{JCC27_1787} decreases significantly the molecule$-$surface
separation distance in the case of the Py and Pz molecules. The
adsorbate-substrate distances are given in {\AA}.}
\label{fig:Geometry2}
\end{figure}
In the next step we performed geometrical relaxations of the molecule$-$metal
systems including the van der Waals forces using the DFT-D approach. Since
the vdW forces are attractive, their effect is to bring the molecule closer to
the surface (see Fig.~\ref{fig:Geometry2}). For the Bz molecule this effect on
the adsorption geometry is not significant, since the average Bz$-$surface
distance is decreased from 2.43 to 2.35\,{\AA}. On the contrary, the inclusion
of the long range dispersion effects has a huge impact on the geometry of the 
Py and Pz molecules adsorbed on Cu(110) surface. The average Py(Pz)--surface
distance decreases significantly from 2.98(2.99) to 2.43(2.59)\,{\AA}. Also 
the tilt angle of the Py molecular plane changes from $+$3$^\circ$ (DFT) to 
$\approx$ $-$7$^\circ$ (DFT$+$D).
\begin{figure*}[htb]
  \begin{tabular}{ccc}
    \multicolumn{3}{c}%
    {\includegraphics[scale=0.35]{DOS-BzPyPz-dft-vdW-003.eps}} \\
    {\includegraphics[scale=0.23]{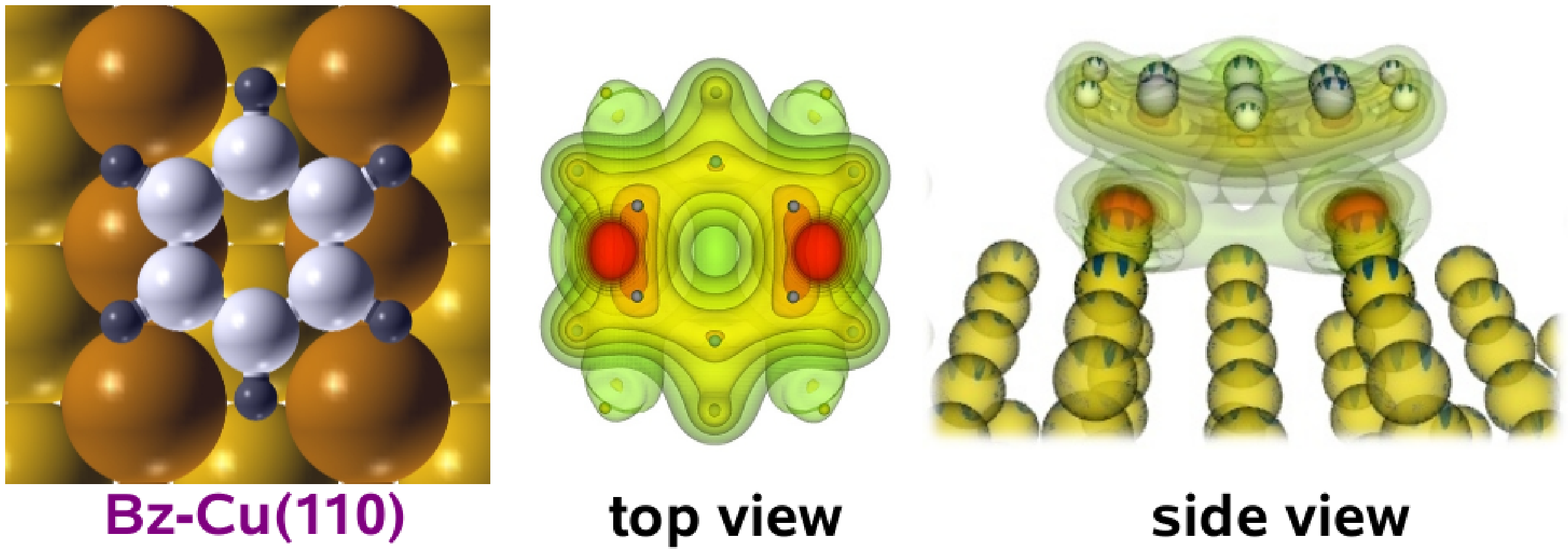}} & 
    {\includegraphics[scale=0.23]{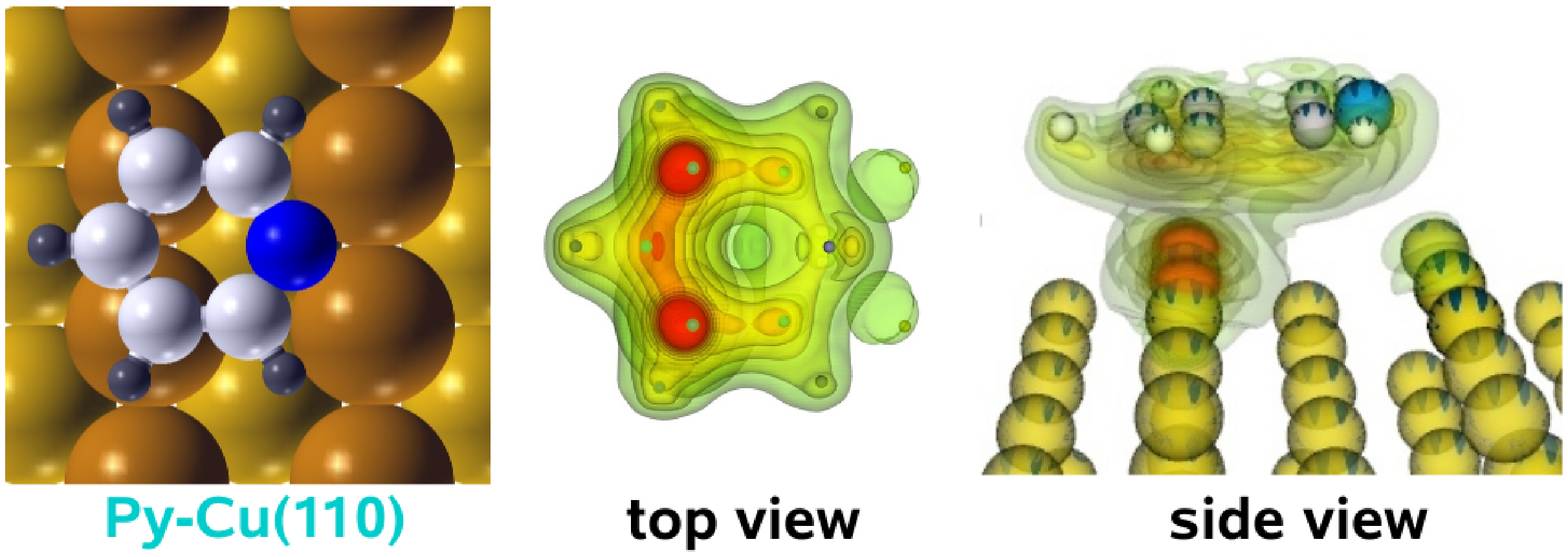}} & 
    {\includegraphics[scale=0.23]{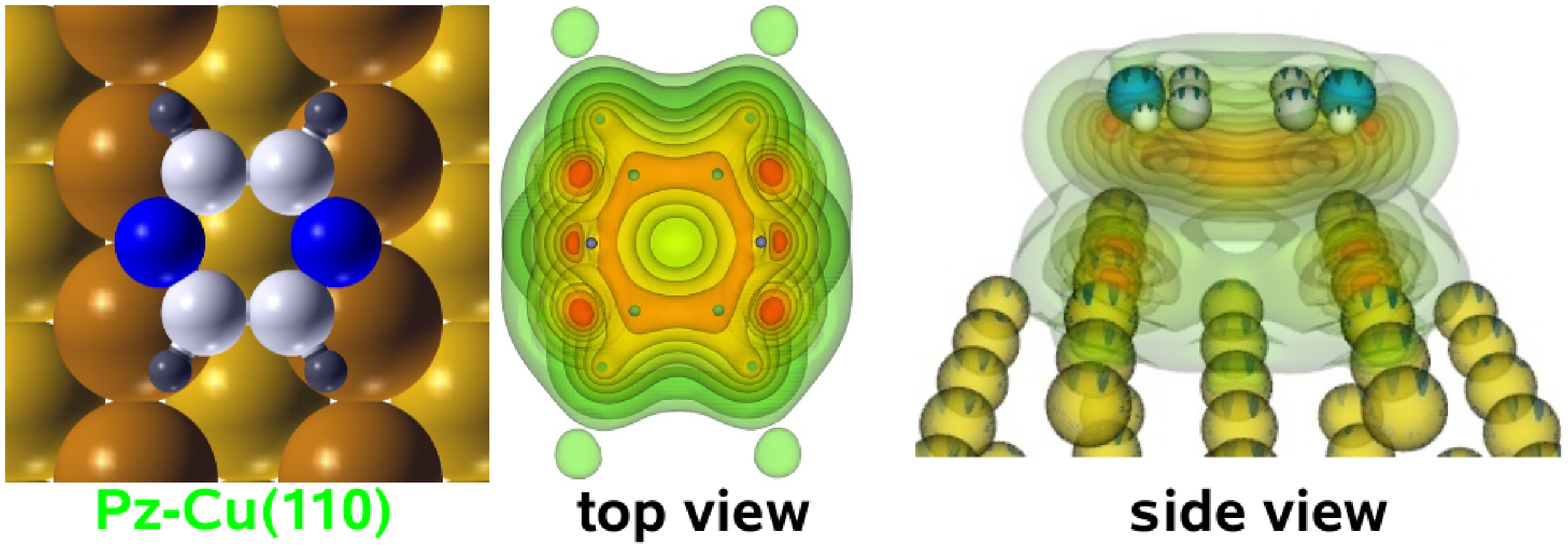}} \\
  \end{tabular}
\caption{(Color online) Upper panel: Local density of states (LDOS) of the 
Bz, Py and Pz molecules adsorbed on Cu(110). In the case of Bz molecule, the
electronic structure of the molecule$-$surface system is not significantly
modified by the inclusion of the dispersion effects. This is not the case of
the Py-- and Pz--Cu(110) systems, where the van der Waals attractive
interaction brings the molecule closer to the surface and allows a stronger
hybridization between the $p_z-$orbitals and the Cu $d-$bands. 
Lower panel: Binding energies due to nonlocal correlation effects evaluated
with the vdW-DF method. Red color indicates a strong non-local (NL)
contribution to the binding energy while green denotes a much weaker
contribution. In the case of the Pz molecule this contribution is delocalized
over the whole heterocycle ring while in the case of the Bz and Py it mainly
arises from the two carbon sites that directly bind to the surface.
}
\label{fig:DosBzPyPzCu110}
\end{figure*}

This structural change drastically affects the \textit{electronic structure}
of the Py-- and Pz--Cu(110) systems. In Fig.~\ref{fig:DosBzPyPzCu110} we
present the local density of states (LDOS) for Bz, Py and Pz molecules in
the case of the relaxed geometries before (DFT) and after inclusion of the van
der Waals forces (DFT$+$vdW). At the molecular site the basic characteristic
displayed by the Bz$-$Cu(110) system are the broad bands with
$\pi_{1,2,3}-$character, while the  $\sigma$ ones are quite sharp and very
localized on the molecule. Also this electronic structure is not affected by
the inclusion of the long-range dispersion effects because the geometry of the
system changes only slightly. On the contrary, for the Py and Pz molecules,
due to a large molecule$-$metal distance in the DFT relaxed geometry, the
$\pi$ bands are less broad being localized mainly on the molecules (see
Fig.~\ref{fig:DosBzPyPzCu110}). The inclusion of the attractive van der Waals 
interaction brings the Py or Pz closer to the surface and thus it allows the
$p_z-$orbitals to hybridize more strongly with the Cu $d-$bands. This
molecule$-$surface interaction results in broad bands with mixed $\pi$ and
metallic character as already seen for the Bz$-$Cu(110) system. 
Note also the appearance in the case of Pz of a broad $\sigma_1$ type band. 

To quantify the importance of the van der Waals interactions in the case of
the flat adsorption process of the Bz, Py and Pz molecules on the Cu(110)
surface, in Table~\ref{tab:EneDiff-001} we report the calculated interaction
$E_{\mathrm{int}}$ and adsorption $E_{\mathrm{ads}}$ energies defined as:
\begin{center}
$E_{\mathrm{int/ads}}=E_{\mathrm{sys}}-
                  (E_{\mathrm{molecule}}^{\mathrm{relaxed/ideal}}+ 
                   E_{\mathrm{Cu(110)}}^{\mathrm{relaxed/clean}})$.
\end{center}

\noindent where $E_{\mathrm{sys}}$ represents the total energy of the relaxed
molecule$-$Cu(110) system, $E_{\mathrm{molecule}}^{\mathrm{relaxed}}$ represents
the total energy of the isolated molecule and
$E_{\mathrm{Cu(110)}}^{\mathrm{relaxed}}$ is the energy of the Cu(110)
surface, both the molecule and surface being in the same atomic configurations
as in the relaxed molecule$-$Cu(110) system. Also, the
$E_{\mathrm{molecule}}^{\mathrm{ideal}}$ denotes the total energy of the
equilibrium isolated molecule and $E_{\mathrm{Cu(110)}}^{\mathrm{clean}}$ is
the energy of the clean Cu(110) surface.

The interaction energy $E_{\mathrm{int}}^{\mathrm{DFT}}$ represents a measure
of the strength of the chemical interaction between molecule and surface, while
the adsorption one $E_{\mathrm{ads}}^{\mathrm{DFT}}$ includes not only the
energy gain due to the bond formation ($E_{\mathrm{int}}^{\mathrm{DFT}}$) but
also the energy paid to deform the molecules and surface from their ideal
configurations to those as found in the relaxed molecule$-$surface system.
\vspace{-0.4cm}
\begin{table}[htb]
\begin{tabular}{c|cc||cccc|}\hline
    & \multicolumn{2}{c}{DFT relaxed}  
    & \multicolumn{4}{c}{DFT$+$vdW relaxed}\\
    &        $E_{\mathrm{int}}^{\mathrm{DFT}}$&
             $E_{\mathrm{ads}}^{\mathrm{DFT}}$& 
             $E_{\mathrm{int}}^{\mathrm{DFT}}$&
             $E_{\mathrm{ads}}^{\mathrm{DFT}}$&
             $E_{\mathrm{ads}}^{\mathrm{vdW-D}}$&
             $E_{\mathrm{ads}}^{\mathrm{vdW-DF}}$
\\\hline
  Bz  &-301&-253&-343&-252&-492&-543 \\
  Py  &-115&-108&-196&-105&-417&-510 \\
  Pz  &-35 & -36& +56& +70&-371&-577 \\\hline
\end{tabular}
\vspace{-0.2cm}
\caption{The interaction and adsorption energies of the Bz, Py 
and Pz molecules adsorbed on the Cu(110) surface (units: meV) for the 
geometries obtained using only DFT (left side) and DFT including the
dispersion corrections (DFT$+$vdW, right side). Note that when the dispersion
effects were considered, the adsorption and interaction energies have been
decomposed into the contribution arising from a
DFT ($E_{\mathrm{ads,int}}^{\mathrm{DFT}}$) and that given by the van der
Waals corrections ($E_{\mathrm{ads,int}}^{\mathrm{vdW-D(F)}}$), i.e., 
$E_{\mathrm{ads,int}}^{\mathrm{DFT+vdW}}=
E_{\mathrm{ads,int}}^{\mathrm{DFT}}+E_{\mathrm{ads,int}}^{\mathrm{vdW-D(F)}}$.
}
\label{tab:EneDiff-001}
\end{table}

\vspace{-0.6cm}
Although the molecular geometries are slightly distorted
from a planar geometry after including the dispersion effects as compared to
the those obtained only with DFT, for Bz and Py the calculated adsorption
energies  $E_{\mathrm{ads}}^{\mathrm{DFT}}$  are \emph{practically invariant}. 
Moreover, the interaction energies $E_{\mathrm{int}}^{\mathrm{DFT}}$ are
lower for the DFT$+$vdW relaxed geometries as compared to those calculated for
the DFT relaxed ones. This theoretical finding implies that by including the
van der Waals dispersions, the \emph{chemical interaction} between the molecule
and surface \emph{increases}! This effect is rather small for Bz since this
molecule is already chemically adsorbed on Cu(110) surface. However, the
inclusion of the vdW interactions lowers significantly the interaction energy
$E_{\mathrm{int}}^{\mathrm{DFT}}$ for Py on Cu(110). Also the average Py-- and
Bz--surface distances are now comparable ($\approx$ 2.43 and 2.35~\AA,
respectively, see also Fig.~\ref{fig:Geometry2}). In consequence, the Py
molecule \textit{becomes chemisorbed} on the Cu(110) surface. In this case,
the the vdW attractive forces are the key ingredient that simply triggers the
chemisorption process of Py on this surface. As regarding the Pz molecule,
the $E_{\mathrm{int}}^{\mathrm{DFT}}$ indicates a weak bonding interaction 
(negative values) for the DFT relaxed geometries which turns into a repulsive 
interaction (positive values) after taking into account the dispersion
effects. However, the adsorption energy due to the van der Waals interaction 
($E_{\mathrm{ads}}^{\mathrm{vdW}}$) is much more negative as compared to 
$E_{\mathrm{int}}^{\mathrm{DFT}}$ such that the total value of the adsorption
energy is negative. This means that the Pz molecule binds to the surface only
through the van der Waals interaction.

In order to deeply understand how the correlation effects contribute to the 
molecule$-$metal interaction, in Fig.~\ref{fig:DosBzPyPzCu110} we plotted the 
binding energies due to the non-local (NL) correlation effects
as calculated within the vdW-DF theory \cite{PRL92_246401}. For the Bz and Py
molecules, the non-local contribution is mainly localized in the regions where
the C atoms are situated on top of Cu atoms of the surface. In the case of Pz,
the contribution of the non-local correlation effects is much more delocalized
over the entire molecule. From the electronic structure point of view, this
peculiar feature is related to a larger contribution to the NL part of the
binding energy of the $\sigma_1$ molecular orbital in the case of the Pz with
respect to the Py molecule. Note that for Bz this contribution is not
significant since the $\sigma_1$ lies deep in energy with respect to the Fermi
energy of the molecule-surface system. Therefore which molecular orbitals will
contribute to the vdW interactions clearly depends on the specific alignment
of these orbitals at the molecule-surface interface. An animation is available
as additional material\cite{epaps}.

To conclude, we prove that the van der Waals dispersion effects together with
the corresponding vdW attractive forces are crucial to reliably calculate in
a self-consistent manner the proper equilibrium adsorption geometry and the
corresponding electronic structure of $\pi-$conjugated heterocycle molecules
adsorbed on Cu(110) surface. The inclusion of the long-range dispersion
effects changes \textit{qualitatively} the adsorption process of Py
(C$_5$H$_5$N) on Cu(110) surface from \emph{physisorption} to
\emph{chemisorption}, while for the Pz (C$_4$H$_4$N$_2$) the van der Waals
interactions represent the driving forces which bind the molecule to the
surface. Our study also clearly pointed out that the vdW interactions depends
on the alignment of the molecular orbitals at adsorbate-substrate interface
and can involve not only $\pi$-like orbitals (perpendicular to the molecular
plane) but also $\sigma$-like orbitals (in the molecular plane). Therefore we
conclude that these effects will definitely play a key role in the adsorption
process of other $\pi-$conjugated heterocycle molecules as porphyrines or
phthalocyanine on metal surfaces.

This work was supported by the DFG (Grants No. SPP1243 and No. HO 2237/3-1), 
Alexander von Humboldt foundation and Japan Society for the Promotion of
Science. The computations were performed at the Forschungszentrum J\"ulich,
Germany.


\begin{thebibliography}{100}
\expandafter\ifx\csname natexlab\endcsname\relax\def\natexlab#1{#1}\fi
\expandafter\ifx\csname bibnamefont\endcsname\relax
  \def\bibnamefont#1{#1}\fi
\expandafter\ifx\csname bibfnamefont\endcsname\relax
  \def\bibfnamefont#1{#1}\fi
\expandafter\ifx\csname citenamefont\endcsname\relax
  \def\citenamefont#1{#1}\fi
\expandafter\ifx\csname url\endcsname\relax
  \def\url#1{\texttt{#1}}\fi
\expandafter\ifx\csname urlprefix\endcsname\relax\def\urlprefix{URL }\fi
\providecommand{\bibinfo}[2]{#2}
\providecommand{\eprint}[2][]{\url{#2}}

\bibitem[{\citenamefont{Tang and VanSlyke}(1987)}]{APL51_913}
\bibinfo{author}{\bibfnamefont{C.~W.} \bibnamefont{Tang}},
  \bibinfo{author}{\bibfnamefont{S.~A.} \bibnamefont{VanSlyke}},
  \bibinfo{journal}{Appl. Phys. Lett.} \textbf{\bibinfo{volume}{51}},
  \bibinfo{pages}{913} (\bibinfo{year}{1987}).

\bibitem[{\citenamefont{(Ed.)}(2004)}]{Shinar_bookOLED}
\bibinfo{author}{\bibfnamefont{J.~S.} \bibnamefont{(Ed.)}},
  \emph{\bibinfo{title}{Organic Light-Emitting Devices}}
  (\bibinfo{publisher}{Springer-Verlag New York, Inc.}, \bibinfo{year}{2004}).

\bibitem[{\citenamefont{Dimitrakopoulos
  et~al.}(1999)\citenamefont{Dimitrakopoulos, Purushothaman, Kymissis,
  Callegari, and Shaw}}]{Science283_822}
\bibinfo{author}{\bibfnamefont{C.~D.} \bibnamefont{Dimitrakopoulos}},
  \bibinfo{author}{\bibfnamefont{S.}~\bibnamefont{Purushothaman}},
  \bibinfo{author}{\bibfnamefont{J.}~\bibnamefont{Kymissis}},
  \bibinfo{author}{\bibfnamefont{A.}~\bibnamefont{Callegari}},
  \bibnamefont{and} \bibinfo{author}{\bibfnamefont{J.~M.} \bibnamefont{Shaw}},
  \bibinfo{journal}{Science} \textbf{\bibinfo{volume}{283}},
  \bibinfo{pages}{822} (\bibinfo{year}{1999}).

\bibitem[{\citenamefont{Dimitrakopoulos and Malenfant}(2002)}]{AdvMat14_99}
\bibinfo{author}{\bibfnamefont{C.}~\bibnamefont{Dimitrakopoulos}},
  \bibinfo{author}{\bibfnamefont{P.}~\bibnamefont{Malenfant}},
  \bibinfo{journal}{Adv. Mater.} \textbf{\bibinfo{volume}{14}},
  \bibinfo{pages}{99} (\bibinfo{year}{2002}).

\bibitem[{\citenamefont{Sundar et~al.}(2004)\citenamefont{Sund
  Podzorov, Menard, Willett, Someya, Gershenson, and Rogers}}]{Science303_1644}
\bibinfo{author}{\bibfnamefont{V.~C.} \bibnamefont{Sundar}},
  \bibinfo{author}{\bibfnamefont{J.}~\bibnamefont{Zaumseil}},
  \bibinfo{author}{\bibfnamefont{V.}~\bibnamefont{Podzorov}},
  \bibinfo{author}{\bibfnamefont{E.}~\bibnamefont{Menard}},
  \bibinfo{author}{\bibfnamefont{R.~L.} \bibnamefont{Willett}},
  \bibinfo{author}{\bibfnamefont{T.}~\bibnamefont{Someya}},
  \bibinfo{author}{\bibfnamefont{M.~E.} \bibnamefont{Gershenson}},
  \bibnamefont{and} \bibinfo{author}{\bibfnamefont{J.~A.}
  \bibnamefont{Rogers}}, \bibinfo{journal}{Science}
  \textbf{\bibinfo{volume}{303}}, \bibinfo{pages}{1644} (\bibinfo{year}{2004}).

\bibitem[{\citenamefont{Green et~al.}(2007)\citenamefont{Green, Choi, Boukai,
  Bunimovich, Johnston-Halperin, DeIonno, Luo1, Sheriff, Xu, Shin
  et~al.}}]{Nature445_414}
\bibinfo{author}{\bibfnamefont{J.~E.} \bibnamefont{Green}},
  \bibinfo{author}{\bibfnamefont{J.~W.} \bibnamefont{Choi}},
  \bibinfo{author}{\bibfnamefont{A.}~\bibnamefont{Boukai}},
  \bibinfo{author}{\bibfnamefont{Y.}~\bibnamefont{Bunimovich}},
  \bibinfo{author}{\bibfnamefont{E.}~\bibnamefont{Johnston-Halperin}},
  \bibinfo{author}{\bibfnamefont{E.}~\bibnamefont{DeIonno}},
  \bibinfo{author}{\bibfnamefont{Y.}~\bibnamefont{Luo1}},
  \bibinfo{author}{\bibfnamefont{B.~A.} \bibnamefont{Sheriff}},
  \bibinfo{author}{\bibfnamefont{K.}~\bibnamefont{Xu}},
  \bibinfo{author}{\bibfnamefont{Y.~S.} \bibnamefont{Shin}},
  \bibnamefont{et~al.}, \bibinfo{journal}{Nature}
  \textbf{\bibinfo{volume}{445}}, \bibinfo{pages}{414} (\bibinfo{year}{2007}).

\bibitem[{\citenamefont{Heimel et~al.}(2006)\citenamefont{Heimel, Romaner,
  Br\'edas, and Zojer}}]{PRL96_196806}
\bibinfo{author}{\bibfnamefont{G.}~\bibnamefont{Heimel}},
  \bibinfo{author}{\bibfnamefont{L.}~\bibnamefont{Romaner}},
  \bibinfo{author}{\bibfnamefont{J.-L.} \bibnamefont{Br\'edas}},
  \bibnamefont{and} \bibinfo{author}{\bibfnamefont{E.}~\bibnamefont{Zojer}},
  \bibinfo{journal}{Phys.~Rev.~Lett.} \textbf{\bibinfo{volume}{96}},
  \bibinfo{pages}{196806} (\bibinfo{year}{2006}).

\bibitem[{\citenamefont{Heimel et~al.}(2007)\citenamefont{Heimel, Romaner,
  Zojer, and Br\'{e}das}}]{NL7_932}
\bibinfo{author}{\bibfnamefont{G.}~\bibnamefont{Heimel}},
  \bibinfo{author}{\bibfnamefont{L.}~\bibnamefont{Romaner}},
  \bibinfo{author}{\bibfnamefont{E.}~\bibnamefont{Zojer}}, 
  \bibinfo{author}{\bibfnamefont{J.-L.} \bibnamefont{Br\'{e}das}},
  \bibinfo{journal}{Nano Lett.} \textbf{\bibinfo{volume}{7}},
  \bibinfo{pages}{932} (\bibinfo{year}{2007}).

\bibitem[{\citenamefont{Atodiresei et~al.}(2008)}]{Atodiresei2008_1}
\bibinfo{author}{\bibfnamefont{N.}~\bibnamefont{Atodiresei}},
\bibinfo{author}{\bibfnamefont{V.}~\bibnamefont{Caciuc}},
\bibinfo{author}{\bibfnamefont{S.}~\bibnamefont{Bl\"ugel}},
\bibinfo{author}{\bibfnamefont{H.}~\bibnamefont{H\"olscher}},
  \bibinfo{journal}{Phys. Rev. B} \textbf{\bibinfo{volume}{77}},
  \bibinfo{pages}{153408} (\bibinfo{year}{2008}).

\bibitem[{\citenamefont{Atodiresei et~al.}(2008)}]{Atodiresei2008_2}
\bibinfo{author}{\bibfnamefont{N.}~\bibnamefont{Atodiresei}},
\bibinfo{author}{\bibfnamefont{V.}~\bibnamefont{Caciuc}},
\bibinfo{author}{\bibfnamefont{H.}~\bibnamefont{H\"olscher}},
\bibinfo{author}{\bibfnamefont{S.}~\bibnamefont{Bl\"ugel}},
  \bibinfo{journal}{Int. J. Quantum Chem.} \textbf{\bibinfo{volume}{108}},
  \bibinfo{pages}{2803} (\bibinfo{year}{2008}).

\bibitem[{\citenamefont{Atodiresei et~al.}(2007)}]{Atodiresei2007_3}
\bibinfo{author}{\bibfnamefont{N.}~\bibnamefont{Atodiresei}},
\bibinfo{author}{\bibfnamefont{K.}~\bibnamefont{Schroeder}},
\bibinfo{author}{\bibfnamefont{S.}~\bibnamefont{Bl\"ugel}},
  \bibinfo{journal}{Phys. Rev. B} \textbf{\bibinfo{volume}{75}},
  \bibinfo{pages}{115407} (\bibinfo{year}{2007}).

\bibitem[{\citenamefont{Atodiresei et~al.}(2007)}]{Atodiresei2007_4}
\bibinfo{author}{\bibfnamefont{N.}~\bibnamefont{Atodiresei}},
\bibinfo{author}{\bibfnamefont{V.}~\bibnamefont{Caciuc}},
\bibinfo{author}{\bibfnamefont{K.}~\bibnamefont{Schroeder}},
\bibinfo{author}{\bibfnamefont{S.}~\bibnamefont{Bl\"ugel}},
  \bibinfo{journal}{Phys. Rev. B} \textbf{\bibinfo{volume}{76}},
  \bibinfo{pages}{115433} (\bibinfo{year}{2007}).

\bibitem[{\citenamefont{Atodiresei et~al.}(2007)}]{Atodiresei2007_5}
\bibinfo{author}{\bibfnamefont{Y.}~\bibnamefont{Mokrousov}},
\bibinfo{author}{\bibfnamefont{N.}~\bibnamefont{Atodiresei}},
\bibinfo{author}{\bibfnamefont{G.}~\bibnamefont{Bihlmayer}},
\bibinfo{author}{\bibfnamefont{S.}~\bibnamefont{Heinze}},
\bibinfo{author}{\bibfnamefont{S.}~\bibnamefont{Bl\"ugel}},
  \bibinfo{journal}{Nanotech.} \textbf{\bibinfo{volume}{18(49)}},
  \bibinfo{pages}{495402} (\bibinfo{year}{2007}).

\bibitem[{\citenamefont{Atodiresei et~al.}(2008)}]{Atodiresei2008_6}
\bibinfo{author}{\bibfnamefont{N.}~\bibnamefont{Atodiresei}},
\bibinfo{author}{\bibfnamefont{P.H.}~\bibnamefont{Dederichs}},
\bibinfo{author}{\bibfnamefont{Y.}~\bibnamefont{Mokrousov}},
\bibinfo{author}{\bibfnamefont{L.}~\bibnamefont{Bergquist}},
\bibinfo{author}{\bibfnamefont{G.}~\bibnamefont{Bihlmayer}},
\bibinfo{author}{\bibfnamefont{S.}~\bibnamefont{Bl\"ugel}},
 \bibinfo{journal}{Phys.~Rev.~Lett.} \textbf{\bibinfo{volume}{100}},
 \bibinfo{pages}{117207} (\bibinfo{year}{2008}).

\bibitem[{\citenamefont{Bruch et~al.}(2007)\citenamefont{Bruch,
Diehl, Venables}}]{Bruch2007}
\bibinfo{author}{\bibfnamefont{L.W.}~\bibnamefont{Bruch}},
  \bibinfo{author}{\bibfnamefont{R.D.}~\bibnamefont{Diehl}},
  \bibinfo{author}{\bibfnamefont{J.A.}~\bibnamefont{Venables}},
 \bibinfo{journal}{Rev.~Modern~Phys.}
  \textbf{\bibinfo{volume}{79}}, \bibinfo{pages}{1381}
  (\bibinfo{year}{2007}).

\bibitem[{\citenamefont{Ooi et~al.}(2006)\citenamefont{Ooi, Rairkar,
  and Adams}}]{Carbon44_231}
\bibinfo{author}{\bibfnamefont{N.}~\bibnamefont{Ooi}},
  \bibinfo{author}{\bibfnamefont{A.}~\bibnamefont{Rairkar}},
   \bibinfo{author}{\bibfnamefont{J.~B.}
  \bibnamefont{Adams}}, \bibinfo{journal}{Carbon}
  \textbf{\bibinfo{volume}{44}}, \bibinfo{pages}{231}
  (\bibinfo{year}{2006}).

\bibitem[{\citenamefont{Ortmann et~al.}(2005)\citenamefont{Ortmann, Schmidt,
  and Bechstedt}}]{PRL95_186101}
\bibinfo{author}{\bibfnamefont{F.}~\bibnamefont{Ortmann}},
  \bibinfo{author}{\bibfnamefont{W.~G.} \bibnamefont{Schmidt}},
  \bibinfo{author}{\bibfnamefont{F.}~\bibnamefont{Bechstedt}},
  \bibinfo{journal}{Phys.~Rev.~Lett.} \textbf{\bibinfo{volume}{95}},
  \bibinfo{pages}{186101} (\bibinfo{year}{2005}).

\bibitem[{\citenamefont{Dion et~al.}(2004)\citenamefont{Dion, Rydberg,
  Schr\"oder, Langreth, and Lundqvist}}]{PRL92_246401}
\bibinfo{author}{\bibfnamefont{M.}~\bibnamefont{Dion}},
  \bibinfo{author}{\bibfnamefont{H.}~\bibnamefont{Rydberg}},
  \bibinfo{author}{\bibfnamefont{E.}~\bibnamefont{Schr\"oder}},
  \bibinfo{author}{\bibfnamefont{D.~C.} \bibnamefont{Langreth}},
  \bibnamefont{and} \bibinfo{author}{\bibfnamefont{B.~I.}
  \bibnamefont{Lundqvist}}, \bibinfo{journal}{Phys.~Rev.~Lett.}
  \textbf{\bibinfo{volume}{92}}, \bibinfo{pages}{246401}
  (\bibinfo{year}{2004}).

\bibitem[{\citenamefont{Chakarova-K\"ack
  et~al.}(2006)\citenamefont{Chakarova-K\"ack, Schr\"oder, Lundqvist, and
  Langreth}}]{PRL96_146107}
\bibinfo{author}{\bibfnamefont{S.~D.} \bibnamefont{Chakarova-K\"ack}},
  \bibinfo{author}{\bibfnamefont{E.}~\bibnamefont{Schr\"oder}},
  \bibinfo{author}{\bibfnamefont{B.~I.} \bibnamefont{Lundqvist}},
  \bibnamefont{and} \bibinfo{author}{\bibfnamefont{D.~C.}
  \bibnamefont{Langreth}}, \bibinfo{journal}{Phys.~Rev.~Lett.}
  \textbf{\bibinfo{volume}{96}}, \bibinfo{pages}{146107}
  (\bibinfo{year}{2006}).

\bibitem[{\citenamefont{Sony et~al.}(2007)\citenamefont{Sony, Puschnig, Nabok,
  and Ambrosch-Draxl}}]{PRL99_176401}
\bibinfo{author}{\bibfnamefont{P.}~\bibnamefont{Sony}},
  \bibinfo{author}{\bibfnamefont{P.}~\bibnamefont{Puschnig}},
  \bibinfo{author}{\bibfnamefont{D.}~\bibnamefont{Nabok}}, \bibnamefont{and}
  \bibinfo{author}{\bibfnamefont{C.}~\bibnamefont{Ambrosch-Draxl}},
  \bibinfo{journal}{Phys.~Rev.~Lett.} \textbf{\bibinfo{volume}{99}},
  \bibinfo{pages}{176401} (\bibinfo{year}{2007}).

\bibitem[{\citenamefont{Ortmann et~al.}(2006)\citenamefont{Ortmann, Bechstedt,
  and Schmidt}}]{PRB73_205101}
\bibinfo{author}{\bibfnamefont{F.}~\bibnamefont{Ortmann}},
  \bibinfo{author}{\bibfnamefont{F.}~\bibnamefont{Bechstedt}},
   \bibinfo{author}{\bibfnamefont{W.~G.}
  \bibnamefont{Schmidt}}, \bibinfo{journal}{Phys.~Rev.~B}
  \textbf{\bibinfo{volume}{73}}, \bibinfo{pages}{205101}
  (\bibinfo{year}{2006}).

\bibitem[{\citenamefont{Grimme}(2006)}]{JCC27_1787}
\bibinfo{author}{\bibfnamefont{S.}~\bibnamefont{Grimme}},
  \bibinfo{journal}{J.~Comput.~Chem.} \textbf{\bibinfo{volume}{27}},
  \bibinfo{pages}{1787} (\bibinfo{year}{2006}).

\bibitem[{\citenamefont{Bl\"ochl}(1994)}]{PRB50_17953}
\bibinfo{author}{\bibfnamefont{P.~E.} \bibnamefont{Bl\"ochl}},
  \bibinfo{journal}{Phys. Rev. B} \textbf{\bibinfo{volume}{50}},
  \bibinfo{pages}{17953} (\bibinfo{year}{1994}).

\bibitem[{\citenamefont{Kresse and Hafner}(1994)}]{Kresse1994}
\bibinfo{author}{\bibfnamefont{G.}~\bibnamefont{Kresse}},
  \bibinfo{author}{\bibfnamefont{J.}~\bibnamefont{Hafner}},
  \bibinfo{journal}{Phys. Rev. B} \textbf{\bibinfo{volume}{49}},
  \bibinfo{pages}{14251} (\bibinfo{year}{1994}).

\bibitem[{\citenamefont{Kresse and Hafner}(1996)}]{Kresse1996}
\bibinfo{author}{\bibfnamefont{G.}~\bibnamefont{Kresse}},
  \bibinfo{author}{\bibfnamefont{J.}~\bibnamefont{Hafner}},
  \bibinfo{journal}{Phys. Rev. B} \textbf{\bibinfo{volume}{54}},
  \bibinfo{pages}{11169} (\bibinfo{year}{1996}).

\bibitem[{\citenamefont{Perdew et~al.}(1996)\citenamefont{Perdew, Burke, and
  Ernzerhof}}]{PRL77_3865}
\bibinfo{author}{\bibfnamefont{J.~P.} \bibnamefont{Perdew}},
  \bibinfo{author}{\bibfnamefont{K.}~\bibnamefont{Burke}},
  \bibinfo{author}{\bibfnamefont{M.}~\bibnamefont{Ernzerhof}},
  \bibinfo{journal}{Phys. Rev. Lett.} \textbf{\bibinfo{volume}{77}},
  \bibinfo{pages}{3865} (\bibinfo{year}{1996}).

\bibitem[{\citenamefont{Makov and M.C.Payne}(1995)}]{Makov1995}
\bibinfo{author}{\bibfnamefont{G.}~\bibnamefont{Makov}},
  \bibinfo{author}{\bibnamefont{M.C.Payne}}, \bibinfo{journal}{Phys. Rev. B}
  \textbf{\bibinfo{volume}{51}}, \bibinfo{pages}{4014} (\bibinfo{year}{1995}).

\bibitem[{\citenamefont{Thonhauser}(2007)}]{Thonhauser2007}
\bibinfo{author}{\bibfnamefont{T.}~\bibnamefont{Thonhauser}},
\bibinfo{author}{\bibfnamefont{V.R.}~\bibnamefont{Cooper}},
\bibinfo{author}{\bibfnamefont{S.}~\bibnamefont{Li}},
\bibinfo{author}{\bibfnamefont{A.}~\bibnamefont{Puzder}},
\bibinfo{author}{\bibfnamefont{P.}~\bibnamefont{Hyldgaard}},
  \bibinfo{author}{\bibfnamefont{D.C.}~\bibnamefont{Langreth}},
  \bibinfo{journal}{Phys. Rev. B} \textbf{\bibinfo{volume}{76}},
  \bibinfo{pages}{125112} (\bibinfo{year}{2007}).

\bibitem[{\citenamefont{Lazi\'{c} et~al.}()\citenamefont{Lazi\'{c}, Atodiresei,
  Alaei, Caciuc, Bl\"ugel, and Brako}}]{JuNoLo}
\bibinfo{author}{\bibfnamefont{P.}~\bibnamefont{Lazi\'{c}}},
  \bibinfo{author}{\bibfnamefont{N.}~\bibnamefont{Atodiresei}},
  \bibinfo{author}{\bibfnamefont{M.}~\bibnamefont{Alaei}},
  \bibinfo{author}{\bibfnamefont{V.}~\bibnamefont{Caciuc}},
  \bibinfo{author}{\bibfnamefont{S.}~\bibnamefont{Bl\"ugel}}, \bibnamefont{and}
  \bibinfo{author}{\bibfnamefont{R.}~\bibnamefont{Brako}},
  \emph{\bibinfo{title}{submitted to {Comp.} {Phys.} {Comm.}} (arXiv:0810.2273v1)}

\bibitem[{\citenamefont{Komeda et~al.}(2004)\citenamefont{Komeda, Kim, Fujita,
  Sainoo, and Kawai}}]{Komeda2004}
\bibinfo{author}{\bibfnamefont{T.}~\bibnamefont{Komeda}},
  \bibinfo{author}{\bibfnamefont{Y.}~\bibnamefont{Kim}},
  \bibinfo{author}{\bibfnamefont{Y.}~\bibnamefont{Fujita}},
  \bibinfo{author}{\bibfnamefont{Y.}~\bibnamefont{Sainoo}},
  \bibinfo{author}{\bibfnamefont{M.}~\bibnamefont{Kawai}}, \bibinfo{journal}{J.
  Chem. Phys.} \textbf{\bibinfo{volume}{120}}, \bibinfo{pages}{5347}
  (\bibinfo{year}{2004}).

\bibitem[{\citenamefont{Rogers et~al.}(2004{\natexlab{a}})\citenamefont{Rogers,
  Shapter, and Ford}}]{Rogers2004}
\bibinfo{author}{\bibfnamefont{B.~L.} \bibnamefont{Rogers}},
  \bibinfo{author}{\bibfnamefont{J.~G.} \bibnamefont{Shapter}},
  \bibnamefont{and} \bibinfo{author}{\bibfnamefont{M.~J.} \bibnamefont{Ford}},
  \bibinfo{journal}{Surf. Sci.} \textbf{\bibinfo{volume}{548}},
  \bibinfo{pages}{29} (\bibinfo{year}{2004}{\natexlab{a}}).

\bibitem[{\citenamefont{Dougherty et~al.}(2006)\citenamefont{Dougherty, Lee,
  and J.~T.~Yates}}]{JPCB110_11991}
\bibinfo{author}{\bibfnamefont{D.~B.} \bibnamefont{Dougherty}},
  \bibinfo{author}{\bibfnamefont{J.}~\bibnamefont{Lee}}, 
  \bibinfo{author}{\bibfnamefont{J.}~\bibnamefont{J.~T.~Yates}},
  \bibinfo{journal}{J.~Phys.~Chem. B} \textbf{\bibinfo{volume}{110}},
  \bibinfo{pages}{11991} (\bibinfo{year}{2006}).

\bibitem[{\citenamefont{Fleming}(1978)}]{Fle78}
\bibinfo{author}{\bibfnamefont{I.}~\bibnamefont{Fleming}},
  \emph{\bibinfo{title}{Frontier Orbitals and Organic Chemical Reactions}}
  (\bibinfo{publisher}{John Wiley $\&$ Sons}, \bibinfo{address}{London},
  \bibinfo{year}{1978}).

\bibitem[{\citenamefont{Zeiss et~al.}(1971)}]{Zeiss1971}
\bibinfo{author}{\bibfnamefont{G. D.}~\bibnamefont{Zeiss}},
  \bibinfo{author}{\bibfnamefont{M. A.}~\bibnamefont{Whitehead}},
  \bibinfo{journal}{J. Chem. Soc. (Inorg. Phys. Theor.)}\textbf{\bibinfo{volume}{A}},
  \bibinfo{pages}{1727} (\bibinfo{year}{1971}).


\bibitem[{\\citenamefont{Atodiresei}(2008)}]{epaps}
\verb|http://www.youtube.com/watch?v=_FqFQVRK4BM| \\
\verb|http://www.youtube.com/watch?v=jeHoDrQht-Q| \\
\verb|http://www.youtube.com/watch?v=pKYIs2nOXFo| \\

\end{thebibliography}
%

\end{document}